\documentclass[english,11pt,a4paper]{article}
\usepackage[style=numeric-comp, sorting=none]{biblatex}
\usepackage{authblk}
\usepackage[text={17.0cm,23.7cm}]{geometry}
\usepackage[T1]{fontenc}
\usepackage{babel}
\usepackage{graphicx}
\usepackage{tabularx}
\usepackage{amsmath}
\usepackage{multirow}
\usepackage{booktabs}
\usepackage{amssymb}
\usepackage[dvipsnames]{xcolor}
\usepackage{yfonts}
\usepackage{ulem}
\usepackage{enumitem}
\usepackage{comment}
\usepackage{mdframed}
\usepackage{float}
\usepackage{mathrsfs}
\usepackage{slashed}
\usepackage[compat=1.1.0]{tikz-feynman}
\usepackage{tikz}
\usepackage{feynmp}
\usepackage{csquotes}
\usepackage[colorlinks=true, allcolors=blue]{hyperref}

\DeclareUnicodeCharacter{202F}{\,}
\DeclareUnicodeCharacter{03BC}{\ensuremath{\mu}}
\DeclareUnicodeCharacter{2212}{-}

\begin{document}

	\title{\vspace{-2cm}

		\vspace{0.6cm}

	\textbf{Direct detection of light dark matter charged under a $L_{\mu}-L_{\tau}$ symmetry}\\[8mm]}
	
	\author[1,2]{Pablo Figueroa\thanks{p.figueroa@campus.lmu.de}}
	\author[3,4,2]{Gonzalo Herrera\thanks{	gonzaloherrera@vt.edu}}
    \author[5]{Fredy Ochoa\thanks{	 faochoap@unal.edu.co}}
    \affil[1]{\normalsize\textit{Ludwig Maximilian Universit\"at M\"unchen, Physics Department, M\"unchen, Germany}}
    \affil[2]{\normalsize\textit{Max-Planck-Institut f\"ur Physik (Werner-Heisenberg-Institut), F\"ohringer Ring 6, 85748 Garching, Germany}}
    \affil[3]{\normalsize\textit{Center for Neutrino Physics, Department of Physics, Virginia Tech, Blacksburg, VA 24061, USA}}
    \affil[4]{\normalsize\textit{Physik-Department, Technische Universit\"at M\"unchen, James-Franck-Stra\ss{}e, 85748 Garching, Germany}}
    \affil[5]{\normalsize\textit{Departamento de Física, Universidad Nacional de Colombia, Ciudad Universitaria, K. 45 No. 26-85, Bogotá D.C., Colombia}}

	\date{}
	
	\maketitle
	
	\begin{abstract}
     A possible extension of the Standard Model able to explain the recent measurement of the anomalous magnetic moment of the muon consists in adding a gauged $U(1)_{L_{\mu}-L_{\tau}}$ symmetry. If the dark matter particle is charged under this symmetry, the kinetic mixing between the new gauge boson and the photon induces dark matter-electron interactions. We derive direct detection constraints on light dark matter charged under a $U(1)_{L_{\mu}-L_{\tau}}$ symmetry with electron recoil experiments, and explore prospects with XLZD and OSCURA to close in the parameter space able to explain simultaneously the recent measurement on the anomalous magnetic moment of the muon and the observed relic density of dark matter. We further discuss the spin-dependent scattering contribution arising in this model, which was ignored previously in the literature.
	\end{abstract}

\section{Introduction}
\label{sec:Intro} 

Some combinations of the family lepton number in the Standard Model may lead to an anomaly free $U(1)$ global charge that can be gauged without the introduction of several new particles. This is the case for the subtraction between pairs of lepton numbers \cite{Foot:1990mn, He:1990pn, He:1991qd}. The corresponding gauge boson of the $U(1)$ symmetry is strongly constrained when the differences involve electron number. However, the $L_{\mu}-L_{\tau}$ model is not so strongly constrained, and there is a region of parameter space able to resolve the apparent discrepancy between the Standard Model prediction for the anomalous magnetic moment of the muon, and the latest experimental measurement at Fermilab \cite{Muong-2:2021ojo,Muong-2:2023cdq}. Another long standing puzzle in the particle physics community is given by the yet unknown particle nature of dark matter \cite{Bertone:2004pz}. In recent years, models where the dark matter particle is lighter than a few GeV, and it is charged under a $U(1)$ extension of the Standard Model have received particular attention, \textit{E.g} \cite{Holdom:1985ag, Galison:1983pa, Arkani-Hamed:2008hhe, Bauer:2018onh, Dutra:2018gmv}. A particular example of these models is a dark matter fermion being charged under a $U(1)_{L_{\mu}-L_{\tau}}$ symmetry, \textcolor{black}{such that its coupling to electrons proceeds via kinetic mixing of the associated gauge boson with the Standard Model hypercharge boson.} 
This model has been studied in the context of GeV-scale Weakly Interacting Massive Particles (WIMPS), \textit{E.g} \cite{Baek:2008nz, Baek:2015fea, Altmannshofer:2016jzy, Biswas:2016yan, Arcadi:2018tly, Bauer:2018egk, Bauer:2018onh}, and in the context of light dark matter, \textit{E.g} \cite{Foldenauer:2018zrz, Choudhury:2020xui,Hapitas:2021ilr,Borah:2021jzu, Deka:2022ltk, AtzoriCorona:2022moj, Manzari:2023gkt, Hooper:2023fqn,Okada_2020, chowdhury2023ultralight}. The model has triggered the interest of the community in recent years, since there is still a region of parameter space able to simultaneously reproduce the observed dark matter relic abundance and the measurement of the anomalous magnetic moment of the muon at Fermilab.

The allowed parameter space able to explain both phenomena is not too large. In this work we will derive updated constraints on this model from dark matter-electron scattering searches at SENSEI-SNOLAB, XENON1T \textcolor{black}{and PANDAX-4T} \cite{SENSEI:2023, Aprile_2019, PandaX:2022xqx}, and we will show that the region able to simultaneously explain the observed dark matter relic abundance and the anomalous moment of the muon may be probed with future experiments such as OSCURA and XLZD \cite{Oscura:2022vmi, Aalbers:2022dzr}. Furthermore, we will calculate the spin-dependent dark matter-electron scattering contribution in this model, showing that it is negligible compared to the spin-independent one.

The paper is organized as follows: In Section \ref{sec:model}, we revisit the dark matter in the $L_{\mu}-L_{\tau}$ model, describing the relevant Lagrangian terms, the estimation of its relic abundance in the early universe within this model, the contribution to the anomalous magnetic moment of the muon, and the spin-independent and spin-dependent scattering cross sections off electrons. In Section \ref{sec:scattering_rates}, we present the formalism relevant to calculate the ionization rates in liquid xenon detectors and semiconductors. In Section \ref{sec:bounds}, we derive updated bounds on the parameter space of the model from recent results of SENSEI-SNOLAB, XENON1T and \textcolor{black}{PANDAX-4T}, and derive projected constraints for future semiconductor and liquid xenon experiments OSCURA and XLZD. Finally, in Section \ref{sec:conclusions}, we present our conclusions.

\section[Dark matter in the mu tau model]{Dark matter in the $U(1)_{L_{\mathbb{\mu}}-L_{\mathbb{\tau}}}$ model}\label{sec:model}

The $U(1)_{L_{\mu}-L_{\tau}}$ extension of the Standard Model plus a fermionic dark matter candidate can be described by the Lagrangian \cite{Foldenauer_2019, Hapitas:2021ilr, Arcadi_2018}

\begin{equation}\label{eq:Lagrangian}
    \mathcal{L} = \mathcal{L}_{L_{\mu}-L_{\tau}} + \mathcal{L}_{\chi},
\end{equation}
where $\mathcal{L}_{L_{\mu}-L_{\tau}}$ refers to the Lagrangian term associated with the $Z'$ gauge boson and $\mathcal{L}_{\chi}$ to the dark matter particle, with

\begin{align}
\begin{split}
        \mathcal{L}_{L_{\mu}-L_{\tau}} = - \frac{1}{4}Z'_{\alpha \beta}Z'^{\alpha \beta} + \frac{1}{2}m^{2}_{Z'}Z'_{\alpha}Z'^{\alpha} + \frac{\epsilon_{0}}{2}Z'_{\alpha \beta}F^{\alpha \beta}
    + g_{\mu \tau}(\bar{\mu}\gamma^{\alpha}\mu - \bar{\tau}\gamma^{\alpha}\tau + \bar{\nu}_{\mu}\gamma^{\alpha}P_{L}\nu_{\mu} - \bar{\nu}_{\tau}\gamma^{\alpha}P_{L}\nu_{\tau})Z'_{\alpha}.
\end{split}
\end{align}
where $F_{\alpha\beta}$ and $Z'_{\alpha\beta}$ corresponds to the photon and $L_{\mu}-L_{\tau}$ strength tensors, respectively. The lepton doublet $L_{\mu} (L_{\tau})$ carries a positive (negative) charge, and $g_{\mu \tau}$ is the gauge coupling between the SM sector and the new gauge boson. The bare kinetic mixing between the field strengths is denoted by $\epsilon_{0}$. The Lagrangian of the dark sector reads

\begin{equation}
    \mathcal{L}_{\chi} = - g_{\chi}\bar{\chi}\gamma_{\mu}\chi Z'^{\mu} - m_{\chi}\bar{\chi} \chi ,
\end{equation}
where \textcolor{black}{$g_{\chi}=g_{\mu\tau}Q_{\chi}$} is a gauge coupling of the dark matter \textcolor{black}{and $Q_{\chi}$ its charge} under the $U(1)_{L_{\mu}-L_{\tau}}$ symmetry group. This model features \textcolor{black}{five} new parameters: $m_{\chi}, m_{Z'},g_{\chi},g_{\mu \tau}$ and $\epsilon_0$ to be constrained and/or determined. Constraints in some of these parameters have been derived in previous works \cite{Foldenauer_2019, Hapitas:2021ilr}, but only in certain scenarios and for fixed relations between $m_{\chi}$ and $m_{Z'}$. Furthermore, the direct detection phenomenology arising from scatterings off electrons was not discussed in detail. In this work we aim to address in a complementary and more extensive way the interplay of these parameters, deriving updated bounds and projections from direct detection experiments sensitive to electron recoils.

Dark matter can acquire the observed relic abundance via the standard freeze out mechanism within this framework. The annihilation cross sections and decay widths that set the relic abundance were calculated in \textit{E.g} \cite{Hapitas:2021ilr, Arcadi_2018}, and here we limit ourselves to show the relation between gauge coupling vs dark matter mass values able to account for the observed relic density of dark matter of the Universe. For sufficiently heavy mediators $m_{Z^{\prime}} \gg m_\chi$, and defining \textcolor{black}{$y \equiv g_\chi^2 g_{\mu \tau}^2\left(m_\chi / m_{Z^{\prime}}\right)^4$}, the cross section and abundance are approximately related via \cite{Kahn_2018}

\begin{equation}
\langle\sigma v\rangle \simeq \frac{3 y}{\pi m_\chi^2} \Longrightarrow \Omega_\chi h^2 \sim 0.1\left(\frac{3 \times 10^{-9}}{y}\right)\left(\frac{m_\chi}{\mathrm{GeV}}\right)^2.
\end{equation}

In the following, we will use this relation to confront the thermal dark matter region of the parameter space with constraints from direct detection. The model also induces a contribution to the muon magnetic moment $a_{\mu}$ due to the $Z'$ exchange at one loop, with value \cite{Hapitas:2021ilr,Baek_2001,Biswas_2017,Gninenko_2001,Patra_2017} 

\begin{equation}
    \delta a_{\mu} = \frac{(g_{\mu \tau}+e \epsilon_{0})^{2}}{4 \pi^{2}} \int_{0}^{1}dz \frac{m^{2}_{\mu}z^{2}(1-z)}{m^{2}_{Z'}(1-z)+m^{2}_{\mu}z^{2}}.
\end{equation}

This shift of the purely SM prediction could be useful to address the $\Delta a_{\mu}$ anomaly \cite{Patra_2017, Andreev_2022}. Here we will confront direct detection results with the combination of $g_{\mu \tau}$ and $m_{Z^{\prime}}$ values yielding the observed anomalous moment of the muon at the Muon g-2 experiment, where the required shift with respect to the Standard Model contribution may be as large as $\delta a_\mu = (24.9 \pm 4.8) \times 10^{-10}$   \cite{Muong-2:2023cdq}.

\begin{figure}[H]
    \centering
    \begin{tikzpicture}[scale=100.0]
    \feynmandiagram[horizontal=a to c,layered layout] {  a --[photon, edge label'=\(\gamma \)] b [dot, label = $\epsilon_{0}$] -- [photon, edge label'=\(Z' \)] c};\end{tikzpicture}\hspace{20mm}
    \begin{tikzpicture}
    \begin{feynman}
    \vertex (a);
    \vertex[right=1.9cm of a] (b);
    \vertex[right=1.9cm of b] (c);
    \vertex[right=1.9cm of c] (d);
    \diagram* {
        (a) -- [photon, edge label'=\(\gamma\)] (b) -- [fermion, half left, looseness=1.5, edge label'=\(\mu/\tau\)] (c) -- [fermion, half left, looseness=1.5] (b),
        (c) -- [boson, edge label'=\(Z'\)] (d),
    };
    \end{feynman}
    \end{tikzpicture}
    \caption{Kinetic mixing between $\gamma$ and $Z'$ boson at leading order and one loop.}
    \label{fig:kineticmixingdiagrams}
\end{figure}
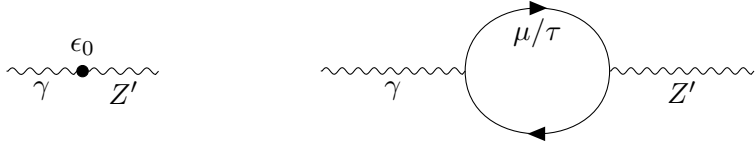

In the following, we calculate the dark matter-electron scattering cross section in the $L_{\mu}-L_{\tau}$ model from the Lagrangian of Eq. \ref{eq:Lagrangian}. The mixing diagrams at tree level and one-loop between the $Z^{\prime}$ and the Standard Model photon are shown in Figure \ref{fig:kineticmixingdiagrams}.

We will factorize the cross section in terms of a spin-independent and a spin-dependent contribution, which was previously neglected in the literature. By following the same parameterization as in \cite{Catena_2020} we can write the invariant matrix element for the dark matter-electron interaction $\mathcal{M}$ as \footnote{The $Z$ exchange diagram between the dark matter and electrons has been neglected, as it is strongly suppressed by a factor $1/m^{2}_{Z}$, and the $Z$ boson is much heavier than the $Z^{\prime}$ masses considered in this work.} (See appendix \ref{sec:appendix} for the full derivation):

%\begin{widetext}
\begin{center}    
\begin{equation*}
    i \mathcal{M} = i \frac{4 m_{\chi}m_{e}g_{\chi}\epsilon e}{(|\vec{q}|^{2} + m^{2}_{Z'})}\Bigg{\{}\delta^{s's}\delta^{r'r} \left( 1 + \frac{|\vec{q}|^{2}}{4m_{\chi}m_{e}}-\frac{|\vec{q}|^{2}}{8\mu^{2}_{\chi e}} - \frac{|\vec{q}|^{2}}{8 \mu_{\chi e}m_{e}}\right)
\end{equation*}
\begin{equation*}
    -i\left( \frac{1}{2m_{\chi}}\delta^{r'r}(\vec{S}^{s's}_{\chi} \times \vec{q}) + \frac{1}{m_{e}}\delta^{s's}(\vec{S}^{r'r}_{e} \times \vec{q})\right) \cdot \vec{v}_{el}^{\perp} \ - \ \frac{1}{m_{\chi}m_{e}}(\vec{S}^{s's}_{\chi} \times \vec{q})\cdot(\vec{S}^{r'r}_{e} \times \vec{q})
\end{equation*}
\begin{equation}\label{matrixelementmutau}
 + \ \left( \frac{\vec{k}}{m_{e}}\right) \cdot \left[ \delta^{s's}\delta^{r'r} \left( -\vec{v}_{el}^{\perp} - \frac{\vec{q}}{4\mu_{\chi e}}\right) + \frac{i}{2}\left( -\frac{1}{m_{e}}\delta^{s's}(\vec{S}^{r'r}_{e} \times \vec{q}) + \frac{1}{m_{\chi}}\delta^{r'r}(\vec{S}^{s's}_{\chi} \times \vec{q})\right)\right] \Bigg{\}}  ,  
\end{equation}
\end{center}
%\end{widetext}
where $\mu_{\chi e}$ is the reduced mass of the dark matter-electron system, the Kronecker deltas $\delta$ are defined in spinor space, the spin operators have been defined as $ 2\vec{S}^{r'r}= \xi^{r'+}\vec{\sigma}\xi^{r}$, $\vec{q}$ is the transferred momentum, $\vec{k}$ is the atomic electron momentum and $\vec{v}_{el}^{\perp}=\vec{v}-\vec{k} / m_e-\vec{q} /\left(2 \mu_{\chi e}\right)$\textcolor{black}{, with $\vec{v}$ the dark matter particle velocity}. It is worth mentioning that Eq. \ref{matrixelementmutau} is calculated via a non relativistic expansion up to first order in $\epsilon$. Furthermore, we neglect the terms proportional to the factor $\Big{(}\frac{\vec{k}}{m_{e}}\Big{)}$ \cite{Catena_2020}. By factorizing out all the pre-factors and splitting the amplitude into a spin-independent $\mathcal{M}_{\rm SI}$ and spin-dependent part  $\mathcal{M}_{\rm SD}$, we get

\begin{equation}
    i \mathcal{M} = i \frac{4 m_{\chi}m_{e}g_{\chi}\epsilon e}{(|\vec{q}|^{2} + m^{2}_{Z'})}\bigg{\{} \mathcal{M}_{\rm SI} + \mathcal{M}_{\rm SD} \bigg{\}},
\end{equation}
where

\begin{equation}
\mathcal{M}_{\rm SI}=\delta^{s^{\prime} s} \delta^{r^{\prime} r}\left(1+\frac{|\vec{q}|^2}{4 m_\chi m_e}-\frac{|\vec{q}|^2}{8 \mu_{\chi e}^2}-\frac{|\vec{q}|^2}{8 \mu_{\chi e} m_e}\right),
\end{equation}
and
\begin{equation}
\mathcal{M}_{\rm SD}=-i\left(\frac{1}{2 m_\chi} \delta^{r^{\prime} r}\left(\vec{S}_\chi^{s^{\prime} s} \times \vec{q}\right)+\frac{1}{m_e} \delta^{s^{\prime} s}\left(\vec{S}_e^{r^{\prime} r} \times \vec{q}\right)\right) \cdot \vec{v}-\frac{1}{m_\chi m_e}\left(\vec{S}_\chi^{s^{\prime} s} \times \vec{q}\right) \cdot\left(\vec{S}_e^{r^{\prime} r} \times \vec{q}\right).
\end{equation}

We can compute independently the two squared matrix elements, by noticing that the cross term $2\Re \{\mathcal{M}_{\rm SI}\mathcal{M}^{*}_{\rm SD}\}$ vanishes when averaging over spins, as the real part is linear in $\vec{S}^{s's}_{\chi}$ and $\vec{S}^{r'r}_{e}$, whose average is zero. From Eq. \ref{matrixelementmutau} it is already clear that the spin-dependent contribution is at least of order $\mathcal{O}(q^2)$, while the spin-independent contribution is of the order 1 plus some minor correction. Concretely, we find for the averaged squared matrix elements:

\begin{equation}
\left|\overline{\mathcal{M}_{\rm SI}}\right|^2=\left[1+\frac{|\vec{q}|^2}{4 m_\chi m_e}-\frac{|\vec{q}|^2}{8 \mu_{\chi e}^2}-\frac{|\vec{q}|^2}{8 \mu_{\chi e} m_e}\right]^2,
\end{equation}
and
\begin{equation}
\left|\overline{\mathcal{M}_{\rm SD}}\right|^2=-\frac{1}{8 m_\chi^2}|\vec{q}|^2|\vec{v}|^2 \sin ^2 \theta-\frac{1}{4 m_e^2}|\vec{q}|^2|\vec{v}|^2 \sin ^2 \theta-\frac{|\vec{q}|^4}{256 m_\chi^2 m_e^2}-\frac{|\vec{q}|^4}{16 m_\chi^2 m_e^2}
\end{equation}
with $\theta$ the scattering angle.

\section{Electron ionization rates in direct detection experiments}\label{sec:scattering_rates}

The dark matter scattering off electrons have been widely studied in recent years, \textit{E.g} \cite{Essig_2012, Essig_2017, Trickle:2019nya, Catena_2020, Liu:2021avx, Catena:2022fnk, Campbell-Deem:2022fqm, Wu:2022jln}. Here we revisit the formalism for the spin-independent and spin-dependent scattering cross sections in the $U(1)_{L_{\mu}-L_{\tau}}$ model. The spin-independent differential ionization cross section in liquid xenon is given by

\begin{equation}
\frac{d\sigma_{\rm ion}^{\rm SI}}{d \ln E_{er}}=\frac{\bar{\sigma}_e}{8 \mu_{\chi e}^2}\int_{q_{\rm min}}^{q_{\rm max}} q d q\left|F_{\rm DM}(\vec{q})\right|^2\left[1+C_{\rm SI}\left(m_\chi, m_e\right)|\vec{q}|^2\right]^2\left|f_{i o n}\left(\vec{k}^{\prime}, \vec{q}\right)\right|^2 \Theta\left(\cos \theta-\frac{v_{\min }}{v}\right),
\end{equation}
where
\begin{equation}
C_{\rm SI}\left(m_\chi, m_e\right)=\left(\frac{1}{4 m_\chi m_e}-\frac{1}{8 \mu_{\chi e}^2}-\frac{1}{8 \mu_{\chi e} m_e}\right).
\end{equation}

$|F_{\rm DM}(q)|$ is known in the literature as the dark matter form factor, introduced to parameterize the momentum transfer dependence of the scattering. In our specific model, there are additional momentum dependent terms, but we factorize the dark matter form factor for reference with other works on dark matter-electron scatterings. It is given by

\begin{equation}\label{eq:DM_formfactor}
    |F_{\rm DM}(\vec{q})|^{2} = \left(\frac{(\alpha m_{e})^{2} + m^{2}_{Z'}}{|\vec{q}|^{2} + m^{2}_{Z'}}\right)^{2} ,
\end{equation}
and we have factorized the dark matter-electron scattering cross section at fixed momentum transfer $q = \alpha m_{e}$ (with $\alpha$ the fine structure constant) at leading order as \cite{Catena_2020}

\begin{equation}\label{sigmae}
    \bar{\sigma}_{e} = \frac{16 \pi \mu^{2}_{\chi e} \alpha \alpha_{\chi} \epsilon^{2}}{(m^{2}_{Z'}+(\alpha m_{e})^{2})^{2}}
\end{equation}
with $\alpha_{\chi} = g^{2}_{\chi}/(4\pi)$, $\mu_{\chi e}$ the reduced mass of the DM-electron system, and $\epsilon$ the kinetic mixing.  Furthermore, $|f_{ion}(\vec{k}´,q)|^{2}$ are the atomic ionization form factors. \textcolor{black}{It is given by the transition probability from the initial (bound electron in the atom with quantum numbers $n,l,m$) to the final state (free electron with quantum numbers $k', l', m'$), concretely \cite{Essig_2012, Catena_2020}
\begin{equation}
f_{ion}(\vec{k},q)=\frac{2 |\vec{k}|^{3}}{(2 \pi)^3}\int \frac{\mathrm{d}^3 \vec{k}}{(2 \pi)^3} \psi_{k^{\prime} \ell^{\prime} m^{\prime}}^{^{\prime}}(\vec{k}+\vec{q}) \psi_{n l m}(\vec{k}).
\end{equation}
}
This factor is independent of the dark matter physics, but its behaviour crucially affect the ionization rates observed at experiments. Finally, we describe the relevant kinematic relations entering in the ionization rates. The minimum dark matter velocity needed to ionize an electron in the $(n,l)$ shell with outgoing energy $E_{er}$ is given by
\begin{equation}
v_{\text {min }}=\frac{E_{e r}+\left|E^{nl}\right|}{|\vec{q}|}+\frac{|\vec{q}|}{2 m_\chi},
\end{equation}
where $E^{nl}$ is the binding energy of the atomic electron. The integration over momentum transfer is performed in the range

\begin{equation*}
q_{\text {min }}=m_\chi v-\sqrt{m_\chi^2 v^2-2 m_\chi(E_{er}+\left|E^{nl}\right|)},
\end{equation*}
\begin{equation}
q_{\max}=m_\chi v+\sqrt{m_\chi^2 v^2-2 m_\chi (E_{e r}+\left|E^{nl}\right|)}.
\end{equation}

We find that the spin-dependent differential ionization cross section is given by

\begin{align}
\frac{d\sigma^{\rm SD}_{\rm ion}}{d \ln E_{er}}= & \frac{\bar{\sigma}_e}{8 \mu_{\chi e}^2}\int_{q_{\rm min}}^{q_{\rm max}} qd q \frac{1}{v}\left|F_{\rm DM}(\vec{q})\right|^2\left\{\left[|\vec{q}|^2|\vec{v}|^2 C_{\rm SD,1}\left(m_\chi, m_e\right)+|\vec{q}|^4 C_{\rm S D,2}\left(m_\chi, m_e\right)\right]\right. \nonumber \\
& \left.-|\vec{q}|^2 v_{\min }^2 C_{\rm SD,1}\left(m_\chi, m_e\right)\right\} \Theta\left(\cos \theta-\frac{v_{m i n}}{v}\right)\left|f_{i o n}\left(\vec{k}^{\prime}, \vec{q}\right)\right|^2,
\end{align}
where we have defined

\begin{equation}
C_{\rm SD,1}=\frac{1}{8 m_\chi^2}+\frac{1}{4 m_e^2} \quad, \quad C_{\rm SD,2}=\frac{1}{256 m_\chi^2 m_e^2}+\frac{1}{16 m_\chi^2 m_e^2}.
\end{equation}

The differential ionization rate is finally given by the convolution of the differential cross section with the incoming dark matter particle flux \cite{Herrera_2021}

\begin{equation}\label{rate}
    \frac{d R_{\rm ion}}{d\ln{E_{er}}} = N_{T} \sum_{n,l} \int d^{3}v \mathcal{F}(\vec{v}+\vec{v}_{e})\frac{d \sigma_{\rm ion}(v,E_{er})}{d \ln{E_{er}}}.
\end{equation}

The dark matter flux on Earth is given by

\begin{equation}
    \mathcal{F}(\vec{v}+\vec{v}_{e})  = \frac{\rho_{\chi}}{m_{\chi}} f(\vec{v}+\vec{v}_{e})
\end{equation}
with $\rho_{\chi} = 0.4$ GeV$/$cm$^{3}$ the local DM density \cite{Read:2014qva,Salucci_2010}, and we assumed a Maxwell-Boltzmann velocity distribution in the detector frame $f(\vec{v}+\vec{v}_{e})$. $N_{T}$ is the number of targets per unit mass and the sum runs over initial quantum numbers of a bound electron in the $(n,l)$ shell of an atom \cite{Essig:2015cda}.

\begin{figure}[H]
    \centering
    \includegraphics[width=0.45\textwidth]{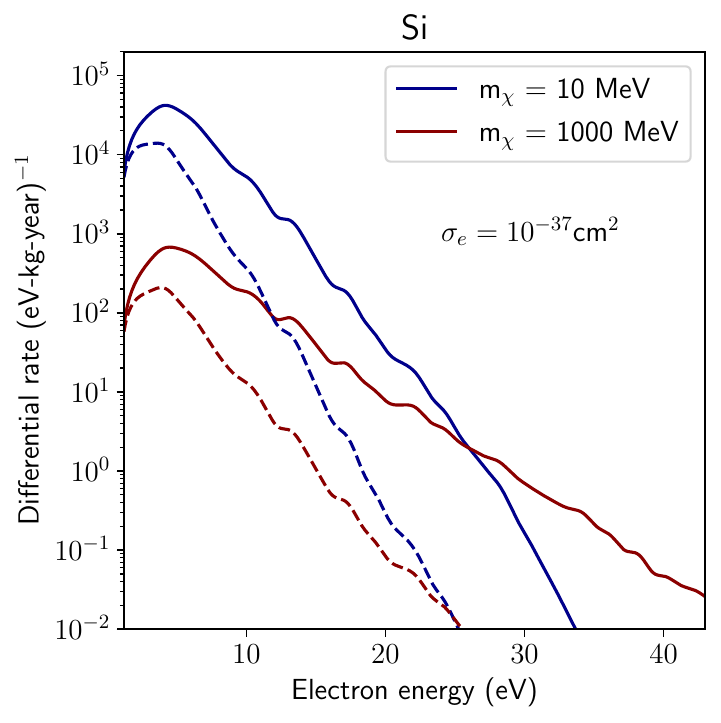}
    \includegraphics[width=0.45\textwidth]{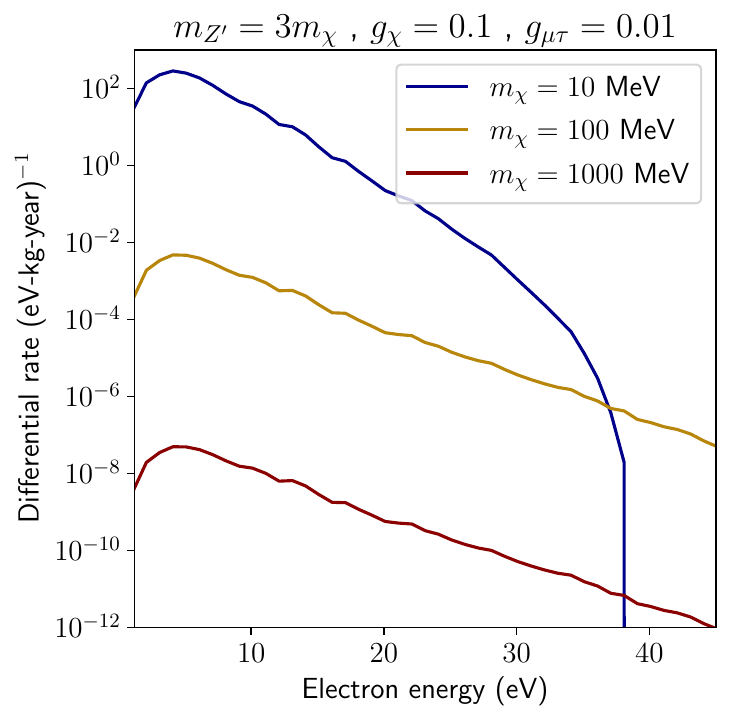}
    \includegraphics[width=0.45\textwidth]{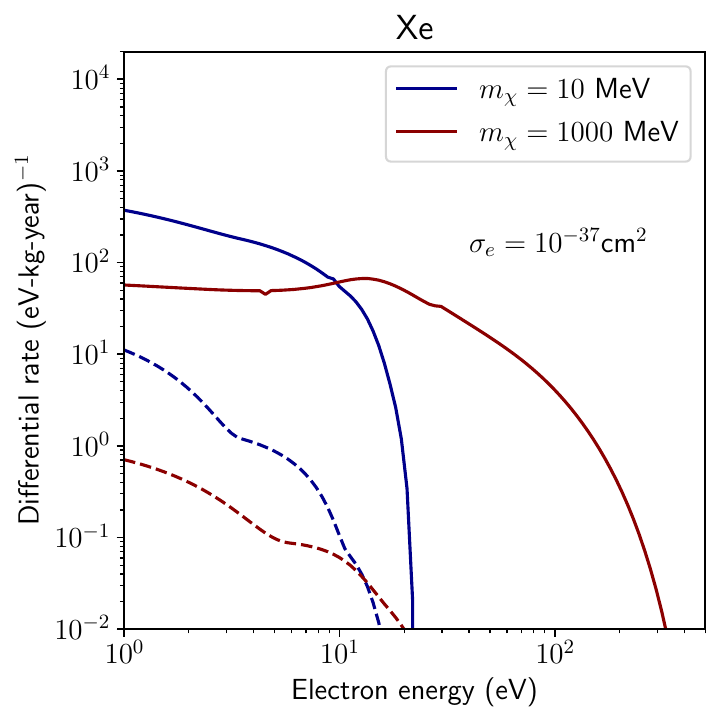}
    \includegraphics[width=0.45\textwidth]{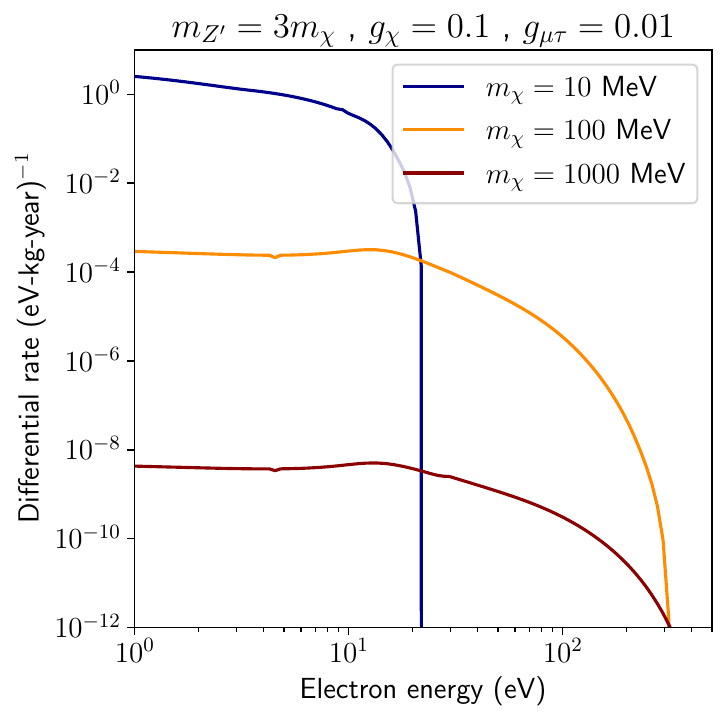}
    \caption{Electron ionization rate induced by dark matter-electron scatterings in Si (upper panels) and Xe (lower panels)  as a function of the electron recoil energy. For the left-side plots, the solid (dashed) lines correspond to a massive (massless) mediator. On the right hand side plots, we show the ionization rate for different values of the dark matter mass, under a fixed relation between the dark matter and mediator mass $m_{Z^{\prime}}=3 m_{\chi}$, fixed dark gauge coupling $g_{\chi}=0.1$, $g_{\mu \tau} = 0.01$ and kinetic mixing with typical value from being generated at one-loop, $\epsilon \simeq g_{\mu \tau}/70$.}
    \label{fig:rates}
\end{figure}

Semiconductors are promising target materials for dark matter-electron interactions because their $\mathcal{O}$(1 eV) band gaps allow ionization signals from dark matter particles as light as a few hundred keV. The scattering rate formalism is similar to the one in liquid xenon, with the main difference that the electron is now part of the band structure in the periodic lattice of a semiconductor crystal. The differential scattering rate to excite an electron from level $i$ to $f$ is given by \cite{Essig:2015cda}

\begin{equation}\label{matrixelement}
      \frac{dR_{cr}}{d\ln{E_{e}}} = \frac{\rho_{\chi}}{m_{\chi}}N_{cell} \alpha \cdot \overline{\sigma}_{e} \frac{m^{2}_{e}}{\mu^{2}_{\chi e}} \int d \ln{q}  \frac{E_{e}}{q} \eta (v_{min})  |F_{\rm DM}(\vec{q})|^{2}|f_{cr}^{i \rightarrow f}(q,E_{e})|^{2},
\end{equation}
where $E_{e}$ the total energy deposited, and $N_{cell} = M_{target}/M_{cell}$ is the number of unit cells in the crystal target. The rate depends on the dark matter velocity distribution via

\begin{equation}
\eta\left(v_{\min }\right)=\int d^3 v f\left(\vec{v}+\vec{v}_e\right) \frac{1}{v} \Theta\left(v-v_{\min}\right).
\end{equation}
where, in the current notation, the minimum velocity necessary for scattering is given by 
\begin{equation}
v_{\min }=\frac{E_{e}}{|\vec{q}|}+\frac{|\vec{q}|}{2 m_\chi}.
\end{equation}

The electronic band structure is contained in the dimensionless crystal form factor, which is an intrinsic property of the target material. It is calculated as \cite{Essig:2015cda}

\begin{equation}\label{cristalfactors}
    f_{cr}^{i \rightarrow f}(\vec{q},\vec{k}) = \sum_{G} \psi_{f}^{*}(\vec{k} + \vec{G} + \vec{q})\psi_{i}(\vec{k}+\vec{G}),
\end{equation}
where the initial and final wave functions of the electrons are described by Bloch functions. Direct detection experiments are not able to measure the deposited energy on the primary electron $E_{e}$, but instead the number of electron-hole pairs produced in an event (ionization signal $Q$) given by \cite{Essig:2015cda}

\begin{equation}
    Q(E_{e}) = 1 + \Bigg{[}\frac{E_{e} - E_{gap}
    }{\varepsilon}\Bigg{]}.
\end{equation}
$E_{gap}$ is the band-gap energy and $\varepsilon$ is the mean energy per electron-hole pair. In addition to the primary electron-hole pair produced by the initial scattering, one extra pair is created for every extra $\varepsilon$ energy deposited above the band gap. In Figure \ref{fig:rates}, we show the electron ionization rate in Silicon (upper plots) and Xenon (lower plots), for different benchmark parameterizations. For the numerical calculation of the scattering rates, we modified the QEDark module \cite{Essig:2015cda} and WIMPrates \cite{Aprile_2019} accordingly. In the left side plots, we show the ionization rates on the very massive mediator (solid) and ultralight or massless mediator (dashed) limits, for fixed values of the dark matter mass and the non-relativistic scattering cross section. On the right side plots, we show the scattering rate for a fixed relation between the dark matter and the mediator masses, $m_{Z^{\prime}}=3m_{\chi}$, dark gauge coupling $g_{\chi}=0.1$, and kinetic mixing given by the irreducible contribution from the loop diagram shown in Figure \ref{fig:kineticmixingdiagrams} \cite{Kamada:2015era,Escudero:2019gzq, Greljo:2021npi, Bauer:2022nwt} 

\begin{equation}
\epsilon=\frac{e g_{\mu \tau}}{12 \pi^2} \log \frac{m_\tau^2}{m_\mu^2} \simeq g_{\mu \tau}/70.
\end{equation}

\begin{figure}[H]
    \centering
    \includegraphics[width=0.49\textwidth]{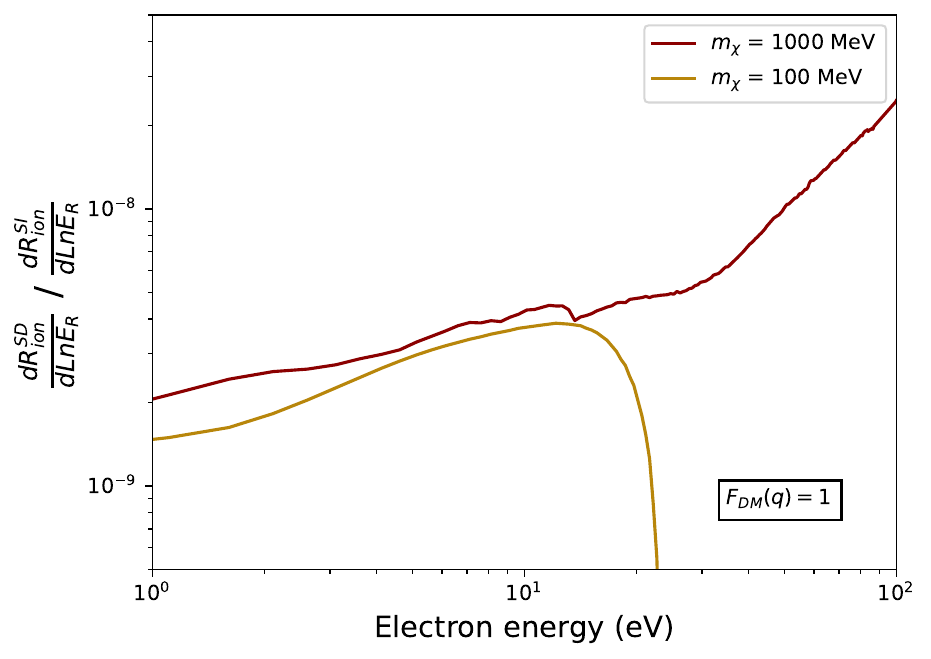}
    \includegraphics[width=0.49\textwidth]{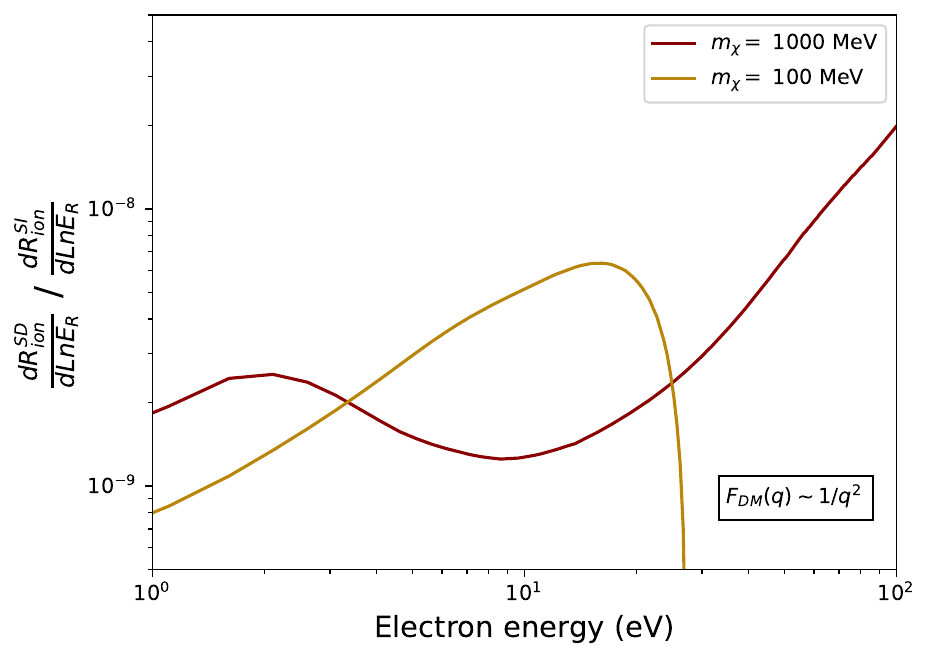}
    \caption{Ratio of ionization rates on liquid xenon due to spin-independent vs spin-dependent dark matter-electron scatterings as a function of the electron recoil energy, for two reference values of the dark matter mass, and a heavy and ultralight/massless mediator, respectively. 
    The corresponding ratios for lower dark matter masses are even smaller than the ones shown in the Figure. 
    }
    \label{fig:ratio}
\end{figure}

In Figure \ref{fig:ratio}, we show the ratio of ionization rates on liquid xenon induced by spin-independent vs spin-dependent dark matter-electron scatterings, as a function of electron energy, for a dark matter mass of $m_{\chi}=100$ MeV (darkyellow) and $m_{\chi}=1$ GeV (red), and for a very heavy mediator (left) and a massless mediator (right). We find that the spin-dependent contribution is suppressed by 7 to 9 orders of magnitude at the relevant energies of direct detection experiments, therefore being negligible compared to the spin-independent contribution. 

\section{Updated constraints and projections}\label{sec:bounds}

In this Section, we present exclusion limits on the parameter space of the $U(1)_{L_{\mu}-L_{\tau}}$ model and the dark matter fermion. The results are presented in various scenarios, linking them to previous studies. For the derivation of the upper limits, we simply impose that the number of events in the region of interest of the experiment shall be lower than the $90\%$ C.L poissonian upper limit on the total number of events reported (or projected) by the experimental collaborations. \textcolor{black}{In particular, we find the $90\%$ C.L upper limit of the number of signal events from each experiment $\varepsilon$ from the poissonian likelihood \cite{Brenner:2022qku}
\begin{equation}
\mathcal{L}\left(N_{\varepsilon}^{\mathrm{sig}}\right)=\frac{\left(N_{\varepsilon}^{\mathrm{sig}}+N_{\varepsilon}^{\mathrm{bck}}\right)^{N_{\varepsilon}^{\mathrm{obs}}}}{N_{\varepsilon}^{\rm {obs }}!} e^{-\left(N_{\varepsilon}^{\mathrm{sig}}+N_{\varepsilon}^{\mathrm{bck}}\right)} .
\end{equation}
where $N_{\varepsilon}^{\mathrm{sig}}$ is the number of signal events, $N_{\varepsilon}^{\mathrm{bck}}$ is the number of background events in the region of interest, and $N_{\varepsilon}^{\mathrm{obs}}$ is the number of observed events. Then, we find the 90\% C.L limit on $N_{\varepsilon}^{\mathrm{sig}}$ after solving $\chi_{\varepsilon}^2-\chi_{\varepsilon, \min }^2 \leq 2.71$, where the $\chi^2$ distribution is given in terms of the experimental likelihood as $\chi_{\varepsilon}^2=-2 \ln \mathcal{L}\left(N_{\varepsilon}^{\text {sig }}\right)$, and $\chi_{\varepsilon, \min }^2$ is the minimum of the distribution.
}

In Table \ref{tab:Experiments} we show the relevant information from all experiments considered in our analysis. \textcolor{black}{The total number of background events $N_{\varepsilon}^{\mathrm{bck}}$ is obtained after integrating the background levels reported by each present or projected experiment over its measured energy range.}

\begin{table}[H]
\begin{center}
\begin{tabular}{|c|c|c|c|c|c|}
\hline 
Experiment &  Material & Exposure & Energy range & Background level  & Events \\
\hline
XENON1T \cite{Aprile_2019} & Xe & 22 t-day & 0.1-4 keV &  1 (keV $\times$ t $\times$ day)$^{-1}$ & 39   \\
\textcolor{black}{PANDAX-4T \cite{PandaX:2022xqx}} & Xe & 0.55 t-year & 0.07-0.23 keV &  10 (keV $\times$ t $\times$ yr)$^{-1}$ & 17   \\
SENSEI-SNOLAB \cite{SENSEI:2023} & Si & 534 g-day & 4.9-16.3 eV & 4.96 (eV $\times$ g $\times$ day)$^{-1}$ & 55  \\
XLZD \cite{Aalbers:2022dzr} & Xe &  200 t-year &  0.1-30 keV &  15.8 (keV $\times$ kg $\times$ year)$^{-1}$ & 39 \\
OSCURA \cite{Oscura:2022vmi} & Si & 30 kg-year &  4.9-16.3 eV & 1 (eV $\times$ kg $\times$ year)$^{-1}$ & 3\\
\hline
\end{tabular}
\end{center}
\caption{Relevant details of the experiments considered in this work.}
\label{tab:Experiments}
\end{table}

\begin{figure}[H]
    \centering
    \includegraphics[width=0.49\textwidth]{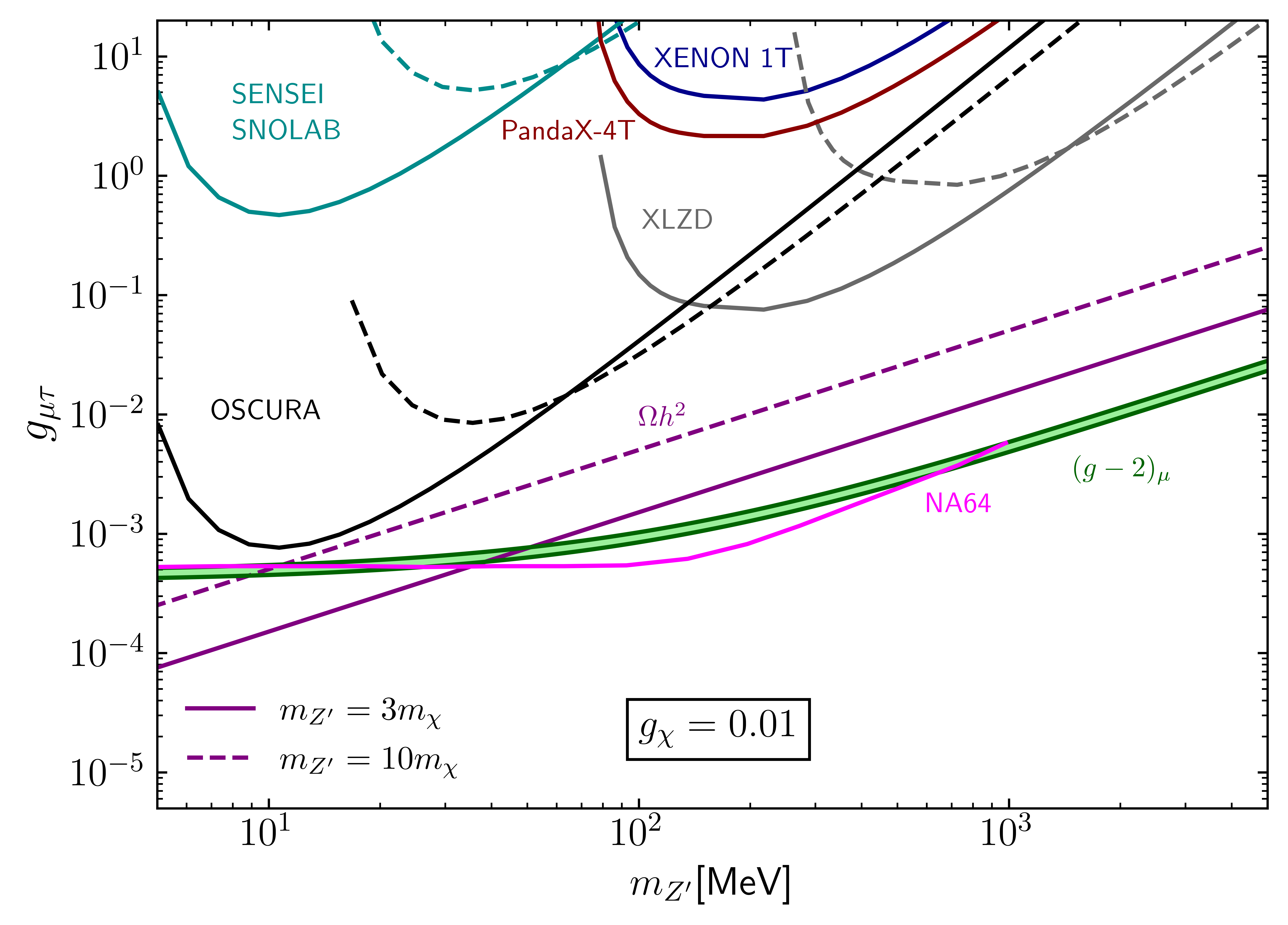}
    \includegraphics[width=0.49\textwidth]{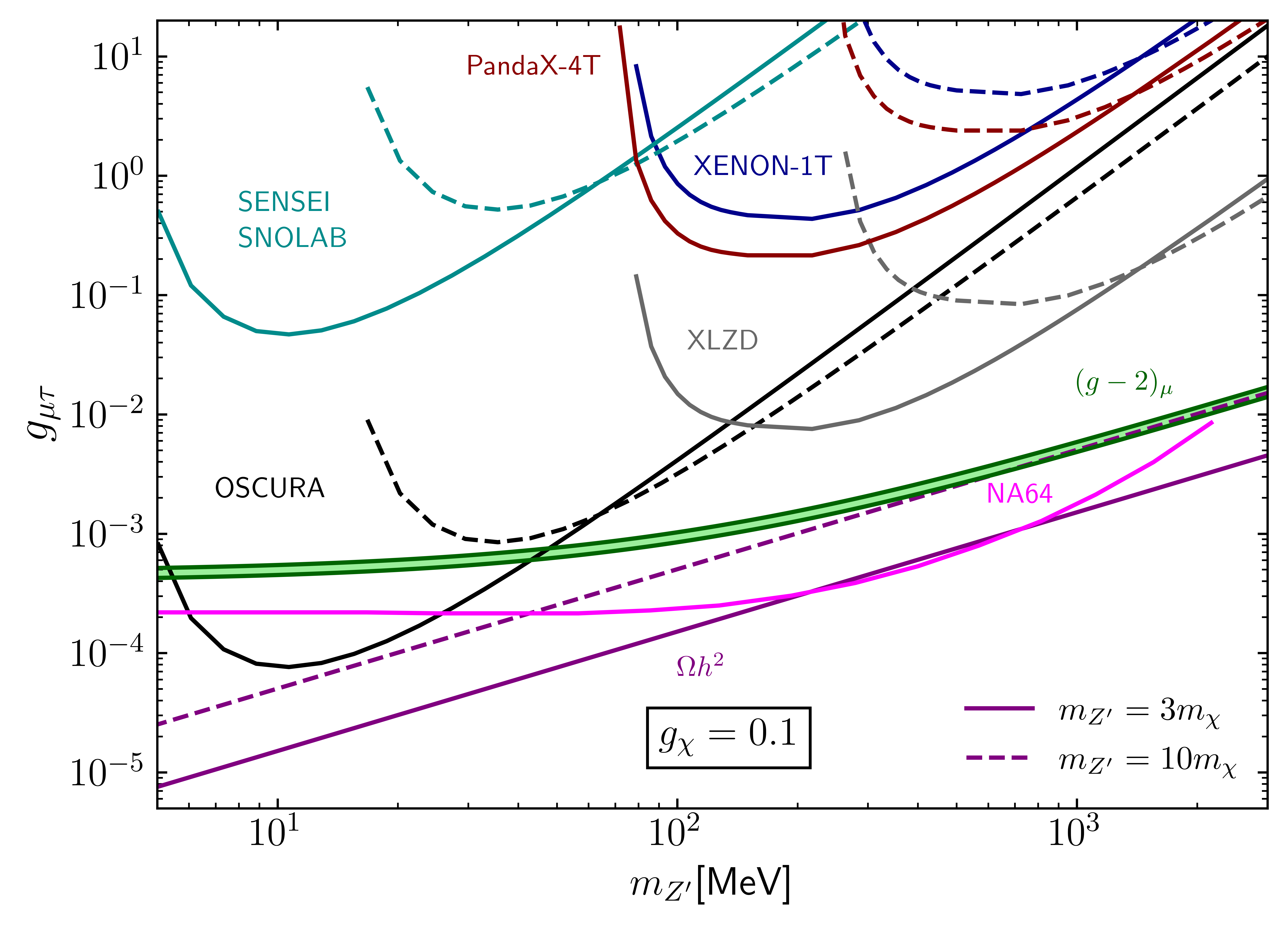}
    \includegraphics[width=0.49\textwidth]{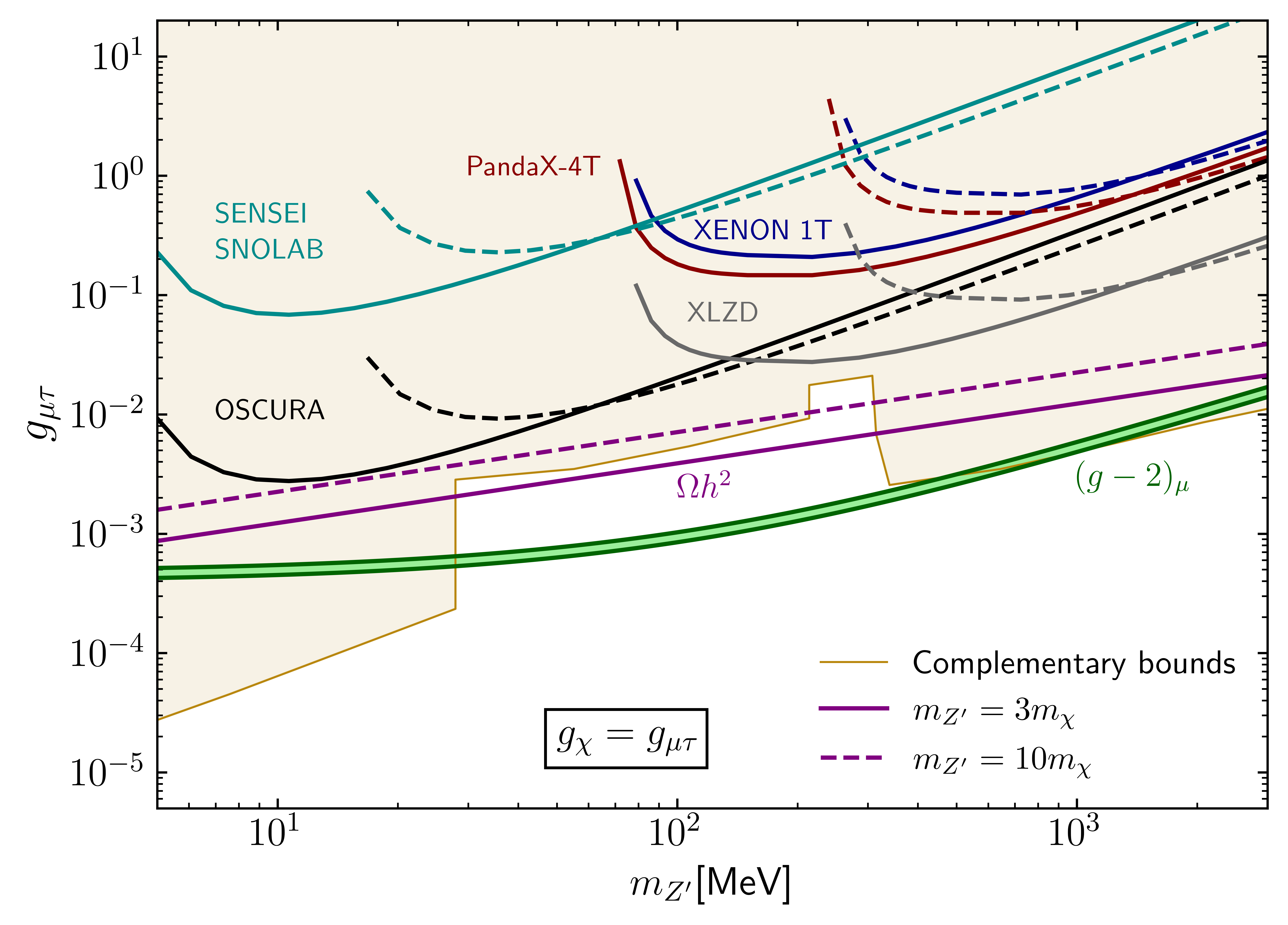}
    \caption{Upper limit on $g_{\mu \tau}$ coupling as function of the mediator mass $m_{Z'}$, for two different values of $m_{\chi}/m_{Z'}$ and considering $g_{\chi} = g_{\mu \tau}$ (lower plot), $g_{\chi} =0.01$ (upper left plot) and $g_{\chi} =0.1$ (upper right plot). We use  $\epsilon \simeq g_{\mu \tau}/70$. \textcolor{black}{The purple solid (dashed) line corresponds to the combination of values able to reproduce the observed relic abundance of dark matter for $m_{Z'}/m_{\chi}=3$ ($m_{Z'}/m_{\chi}=10$). We rescaled the results from \cite{Hapitas:2021ilr, Biswas:2016yan}}. The green band corresponds to the combination of values able to explain the $(g-2)_{\mu}$ anomaly.}
    \label{fig:UL_xenon_sensei}
\end{figure}

In Figure \ref{fig:UL_xenon_sensei}, we show 90 $\%$ C.L upper limits from SENSEI-SNOLAB (cyan), XENON1T (blue), OSCURA (black) and XLZD (grey) on the gauge coupling $g_{\mu \tau}$ vs the $Z^{\prime}$ mediator mass, assuming different relations between $g_{\mu \tau}$ and $g_{\chi}$, and two relations between the mediator and dark matter masses, $m_{Z^{\prime}}=3m_{\chi}$ (solid) and  $m_{Z^{\prime}}=10 m_{\chi}$ (dashed). For comparison, we show in orange complementary constraints from measurements of the effective number of cosmological neutrinos $N_{\rm eff}$ \cite{Escudero:2019gzq}, CMB \cite{Padmanabhan:2005es, Slatyer:2012yq, Asai:2020qlp}, neutrino trident production \cite{CHARM-II:1990dvf,Altmannshofer_2014} and colliders \cite{Foldenauer:2018zrz}. Additional astrophysical constraints arise from cosmic ray cooling in Active Galactic Nuclei \cite{Herrera:2023nww} and cosmic ray electron boosted dark matter \cite{Ema:2018bih, Granelli:2022ysi}. The recent results of NA64 have been re-scaled accordingly \cite{NA64:2024klw}.  \textcolor{black}{Concretely, in the upper left panel of Figure \ref{fig:UL_xenon_sensei}, we display in magenta color the limit obtained by the collaboration on the gauge boson, via missing energy-momentum technique. In the upper right panel of the Figure \ref{fig:UL_xenon_sensei}, we display the more restrictive limit obtained from decays of the gauge boson into dark matter particles, assuming $g_{\chi}=0.1$ and $m_{Z^{\prime}}=3m_{\chi}$.}  We also confront our bounds with the region of the parameter space able to explain the relic density of dark matter (solid and dashed purple) \cite{Foldenauer:2018zrz, Hapitas:2021ilr}, and the anomalous magnetic moment of the muon (green) \cite{Muong-2:2023cdq}.

As can be noticed in the plots, thermal light dark matter in this model with $g_{\mu \tau}=g_{\chi}$ is two orders of magnitude below current constraints (SENSEI-SNOLAB, XENON1T and \textcolor{black}{PANDAX-4T}), and it may be \textcolor{black}{close to be} partially probed for future experiments like OSCURA and XLZD. Furthermore, the combination of parameters able to explain the recent measurement on the anomalous moment of the muon and the observed relic density of dark matter could be probed (XLZD and OSCURA) for certain choices of the dark gauge coupling $g_{\chi}$. This plot clearly shows the discovery potential of muonic forces with low-threshold dark matter detectors. It should be noted that the new gauge boson is constrained by colliders and beam dump experiments at masses above $m_{Z^{\prime}} \gtrsim 0.5$ GeV \cite{Foldenauer:2018zrz}, so the most interesting region of parameter space lies in between this value and the cosmological bound. There are additional constraints on the new mediator from neutrino-electron and neutrino-nucleus scatterings in this model, however, we find these to be unable to probe the region of parameter space favored by the $g-2$ muon anomaly \cite{Harnik:2012ni}. Solar neutrinos at direct detection experiments may allow to probe new mediators, but not the dark matter particle nature. However, they may constitute a Standard Model background for the dark matter detection via ionization signatures \cite{Essig:2018tss, Herrera:2023xun, Carew:2023qrj}, and also a background for dark matter in the context of the $L_{\mu}-L_{\tau}$ model, \textit{E.g} \cite{Harnik:2012ni, Cerdeno:2016sfi, Amaral:2021rzw, Li:2022jfl, DeRomeri:2024dbv}.

\begin{figure}[H]
    \centering
    \includegraphics[width=0.45\textwidth]{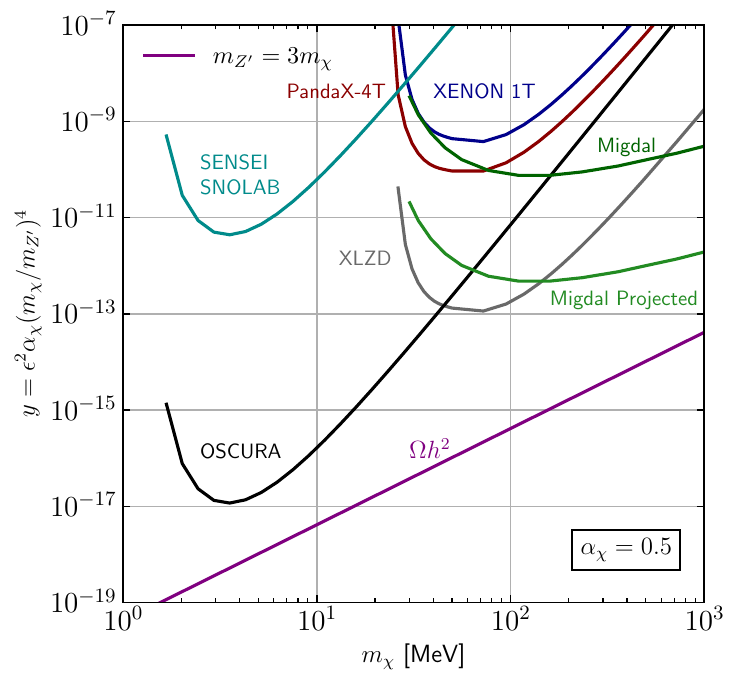}
    \caption{\textcolor{black}{Upper limits on the parameter $y= \epsilon^2 \alpha_{\chi}\left(m_\chi / m_{Z^{\prime}}\right)^4$, from a variety of electron recoil experiments. For comparison, we show current limits from the Migdal ionization signal induced by nuclear recoils in PandaX-4T \cite{PandaX:2022xqx}, and projected Migdal ionization limits rescaling the exposure to that projected by the XLZD experiment. Furthermore, we show the thermal target for this fixed choice of ration between mediator and dark matter masses, $m_{Z'}/m_{\chi}=3$.}}
    \label{fig:limits_y}
\end{figure}

\textcolor{black}{In Figure \ref{fig:limits_y}, we display limits on the parameter $y= g_\chi^2 g_{\mu \tau}^2\left(m_\chi / m_{Z^{\prime}}\right)^4$, from all experiments discussed previously, and compare with limits derived from the Migdal ionization signal from nuclear recoils in PandaX-4T \cite{PandaX:2022xqx}. Furthemore, we show estimated projected limits on this parameter space from the Migdal effect in the future XLZD experiment. For this purpose, we simply rescale the limits from PandaX-4T with the projected exposure of XLZD. We have checked that the energy threshold, experimental resolution and efficiency functions in PandaX-4T and the XENON1T experiment have an effect on the limits of less than $10\%$ for the dark matter masses of interest, but our projected limit should still be regarded as estimative. In the Figure, we also show the thermal relic target in such parameter space, to allow for comparison with experimental limits. For various masses, the projected limits from OSCURA and XLZD lie remarkably close to such thermal values, which indicates that these experiments may be able to probe certain regions even for this relation of mediator and dark matter masses $m_{Z^{\prime}}/m_{\chi}=3$. The prospects are more promising for larger ratios of mediator and dark matter mass, as we will show in the following.}

\begin{figure}[H]
    \centering
    \includegraphics[width=0.45\textwidth]{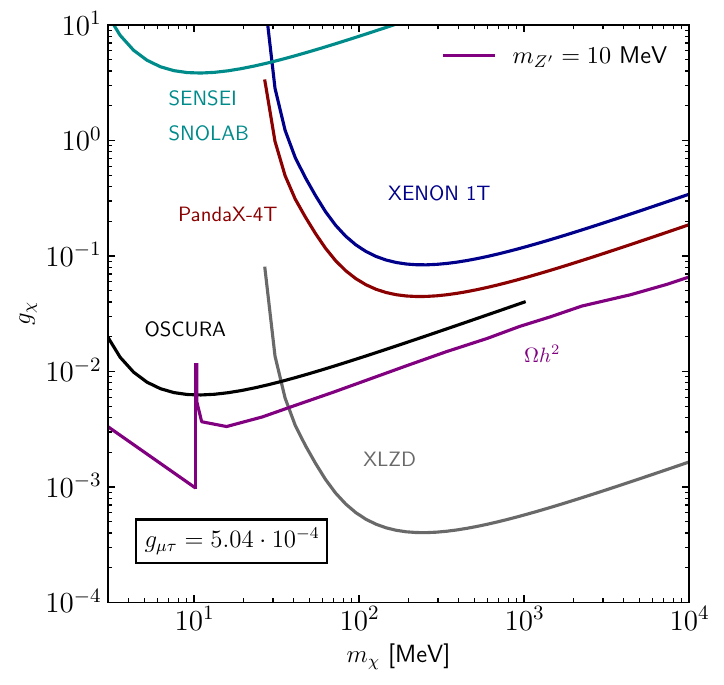}
    \includegraphics[width=0.45\textwidth]{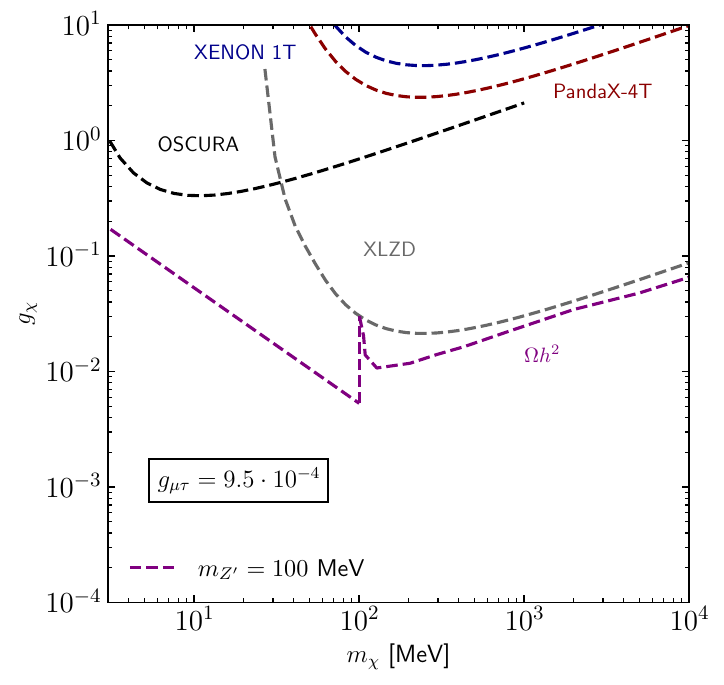}
    \caption{Upper limits on $g_{\chi}$ as a function of $m_{\chi}$, for fixed values of $g_{\mu \tau}$ able to explain the anomalous magnetic moment of the muon. Thermal dark matter able to explain the anomalous magnetic moment of the muon could be probed by future experiments XLZD and OSCURA in a wide range of parameter space.}
    \label{fig:UL_xenon_sensei_mchi}
\end{figure}

\begin{figure}[H]
    \centering
    \includegraphics[width=0.45\textwidth]{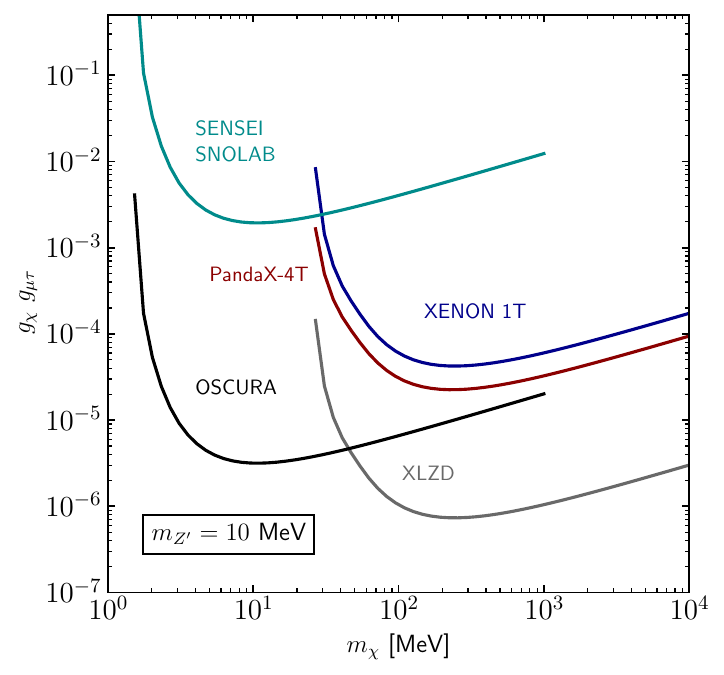}
    \includegraphics[width=0.45\textwidth]{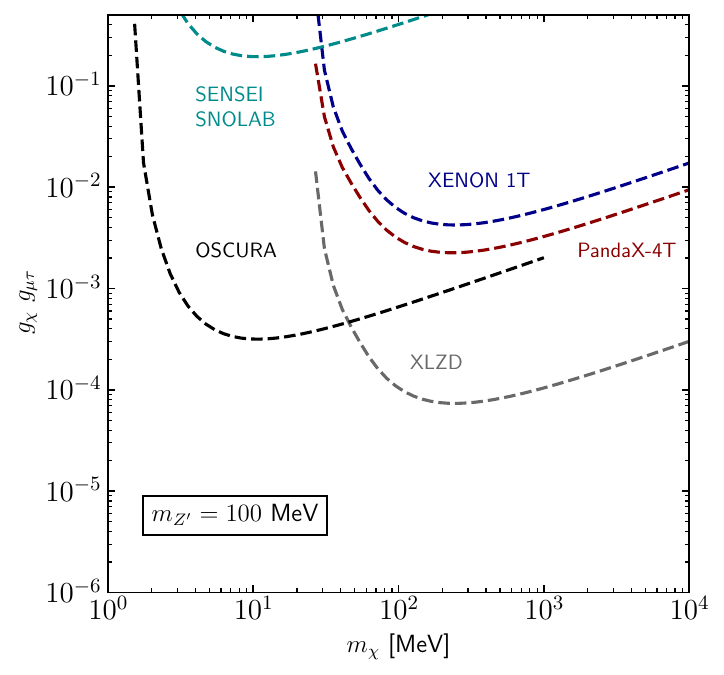}
    \caption{Upper limits on $g_{\chi}  g_{\mu \tau}$ as a function of $m_{\chi}$, for fixed values of $m_{Z'}$.}
    \label{fig:UL_gxgmu_xenon_sensei_mchi}
\end{figure}

In Figure \ref{fig:UL_xenon_sensei_mchi}, we show upper limits in the parameter space of dark gauge coupling $g_{\chi}$ vs dark matter mass $m_{\chi}$, for fixed values of the gauge coupling $g_{\mu \tau}$ able to explain the muon $g-2$ anomaly, and two values of the mediator mass $m_{Z^{\prime}}=10$ MeV ($m_{Z^{\prime}}=100$ MeV), for which $g_{\mu \tau}=5.04 \times 10^{-4}$ ($g_{\mu \tau}= 9.5 \times 10^{-4}$). As can be appreciated in the Figure, current constraints from XENON1T \textcolor{black}{and PandaX-4T} lie less than an order magnitude away from the thermal prediction, and future experiments OSCURA and XLZD may be able to close in a large portion of parameter space for mediator masses $m_{Z^{\prime}} \lesssim 100$ MeV. This indicates that dark matter charged under a $L_{\mu}-L_{\tau}$ may be strongly constrained at the MeV scale in the near future via the complementarity between two different observables: measurements of anomalous magnetic moment of the muon and ionization rates in direct detection experiments. For completeness, we also show in Figure \ref{fig:UL_gxgmu_xenon_sensei_mchi} the upper limits for the product $g_{\chi} g_{\mu \tau}$.

\section{Conclusions}\label{sec:conclusions}
We have derived direct detection constraints on dark matter charged under a $U(1)_{L_{\mu}-L_{\tau}}$ symmetry, via its scatterings off electrons in current and future low-energy liquid xenon and semiconductor detectors. In particular, we have derived constraints from XENON1T, \textcolor{black}{PANDAX-4T}, SENSEI-SNOLAB, XLZD and OSCURA. Moreover, we have calculated the spin-dependent scattering contribution arising in this model, and we have shown that it is generally negligible compared to the spin-independent one at the energy scales of direct detection searches.

Current sensitivity from SENSEI-SNOLAB, XENON1T and \textcolor{black}{PANDAX-4T} isn't sufficiently strong to constrain the theoretically expected values from thermal dark matter production in the $L_{\mu}-L_{\tau}$ model at the MeV scale. However, projected constraints from OSCURA and XLZD will allow to probe a substantial region of the parameter space of dark matter couplings and masses able to simultaneously explain the observed dark matter relic abundance and the anomalous magnetic moment of the muon. Only if the gauge boson mediating the interaction between dark matter and electrons is much heavier than the dark matter particle ($m_{Z^{\prime}} \gtrsim 100$ MeV), direct detection experiments would not be able to probe a substantial part of parameter space accounting for the relic density of dark matter.

We hope that incoming direct detection experiments will allow to robustly test light dark matter charged under a $L_{\mu}-L_{\tau}$ symmetry at the MeV scale. In this work, we have shown that this task would become feasible with future experiments like OSCURA and XLZD.

\subsection*{Acknowledgments}
GH is grateful to Patrick Huber for discussions on the future XLZD experiment.  PF is grateful to Vanessa Zema for discussions on the semiconductor experiments. The work of GH is supported by the U.S. Department of Energy Office of Science under award number DE-SC0020262, by the Collaborative Research Center SFB1258, and by the Deutsche Forschungsgemeinschaft (DFG, German Research Foundation) under Germany's Excellence Strategy - EXC-2094 - 390783311. F. O. Thanks the support from Science Faculty of Universidad Nacional de Colombia under the internal project with HERMES code 56492.

\appendix

\section{Derivation of matrix element for elastic dark matter-electron scattering }\label{sec:appendix}

In this work we have focused in the regime where  $m_{Z^{\prime}} \ll  m_Z$. Besides that, we have focused on dark matter masses larger than the electron mass, whose velocity is typically greater than the dark matter particle velocity \cite{Catena_2020}. The $Z$-boson contribution is suppressed by a factor $m_{Z}^4$ in the light regime, so that it can be ignored and the amplitude is finally written as (up to first order in $\epsilon$)

\begin{equation}
i \mathcal{M}_{Z^{\prime}}=-i g_\chi \epsilon e \frac{1}{\left(|\vec{q}|^2+m_{Z^{\prime}}^2\right)}\left[\bar{u}^{s^{\prime}}\left(p^{\prime}\right) \gamma_\mu u^s(p)\right]\left[\bar{u}^{r^{\prime}}\left(k^{\prime}\right) \gamma^\mu u^r(k)\right].
\end{equation}

For the spinors we adopt the representation
$$
\left(\begin{array}{c}
u_L \\
u_R
\end{array}\right)=\left(\begin{array}{cc}
\sqrt{E-\vec{p} \cdot \vec{\sigma}} \hspace{2mm} \xi^s \\
\sqrt{E+\vec{p} \cdot \vec{\sigma}} \hspace{2mm} \xi^s
\end{array}\right)
$$
and
$$
u^s(p)=\frac{1}{\sqrt{2}}\left(\begin{array}{l}
u_R+u_L \\
u_R-u_L
\end{array}\right)
$$
In the non-relativistic limit the spinors can be written as:

\begin{equation}
%\begin{aligned}
    u^r(k) = \sqrt{2 m_e}
    \begin{pmatrix}
        \xi^r \\
        \frac{\vec{k} \cdot \vec{\sigma}}{2 m_e} \xi^r
    \end{pmatrix} \hspace{5mm} , \hspace{5mm}
    u^s(p) = \sqrt{2 m_\chi}
    \begin{pmatrix}
        \xi^s \\
        \frac{\vec{p} \cdot \vec{\sigma}}{2 m_\chi} \xi^s
    \end{pmatrix}
%\end{aligned}
\end{equation}

\noindent where the $\gamma$ matrices in the Dirac representation are given by:
$$
\gamma^0=\left(\begin{array}{cc}
1 & 0 \\
0 & -1
\end{array}\right) \quad, \quad \gamma^i=\left(\begin{array}{cc}
0 & \sigma^i \\
-\sigma^i & 0
\end{array}\right) \quad, \quad \gamma^5=\left(\begin{array}{ll}
0 & 1 \\
1 & 0
\end{array}\right)
$$
After dealing with the different spinor structures

\begin{equation}
\begin{gathered}
i \mathcal{M}=i \frac{4 m_\chi m_e g_\chi \epsilon e}{\left(|\vec{q}|^2+m_{Z^{\prime}}^2\right)}\left\{\xi^{s^{\prime} \dagger} \xi^s \xi^{r^{\prime} \dagger} \xi^r+\frac{1}{4 m_e^2} \xi^{s^{\prime} \dagger} \xi^s \xi^{r^{\prime} \dagger}(\overrightarrow{k^{\prime}} \cdot \vec{\sigma})(\vec{k} \cdot \vec{\sigma}) \xi^r+\frac{1}{4 m_\chi^2} \xi^{r^{\prime}\dagger} \xi^r \xi^{s^{\prime} \dagger}(\overrightarrow{p^{\prime}} \cdot \vec{\sigma})(\vec{p} \cdot \vec{\sigma}) \xi^s\right. \\
\quad-\frac{1}{4 m_\chi m_e}\left[\xi^{s^{\prime} \dagger} \sigma_i(\vec{p} \cdot \vec{\sigma}) \xi^s \xi^{r^{\prime} \dagger} \sigma^i(\vec{k} \cdot \vec{\sigma}) \xi^r+\xi^{s^{\prime}\dagger} \sigma_i(\vec{p} \cdot \vec{\sigma}) \xi^s \xi^{r^{\prime}\dagger}(\overrightarrow{k^{\prime}} \cdot \vec{\sigma}) \sigma^i \xi^r\right. \\
\left.\left.\quad+\xi^{s^{\prime} \dagger}(\overrightarrow{p^{\prime}} \cdot \vec{\sigma}) \sigma_i \xi^s \xi^{r^{\prime} \dagger} \sigma^i(\vec{k} \cdot \vec{\sigma}) \xi^r+\xi^{s^{\prime}\dagger}(\overrightarrow{p^{\prime}} \cdot \vec{\sigma}) \sigma_i \xi^s \xi^{r^{\prime} \dagger}(\overrightarrow{k^{\prime}} \cdot \vec{\sigma}) \sigma^i \xi^r\right]\right\}
\end{gathered}
\end{equation}

Introducing the value of the momentum transfer and the variable $\vec{v}_{el}^{\perp}$ that comes from the energy conservation
\begin{equation}
\vec{q}=\vec{p}-\vec{p}^{\prime}=\vec{k}^{\prime}-\vec{k} \quad, \quad \vec{v}_{e l}^{\perp}=\vec{v}-\frac{\vec{q}}{2 \mu_{\chi e}}-\frac{\vec{k}}{m_e},
\end{equation}
and defining the spin operator as
\begin{equation}
2 \vec{S}^{r^{\prime} r}=\xi^{r^{\prime} \dagger} \vec{\sigma} \xi^r,
\end{equation}
the amplitude can be finally written as
$$
\begin{gathered}
i \mathcal{M}=i \frac{4 m_\chi m_e g_\chi \epsilon e}{\left(|\vec{q}|^2+m_{Z^{\prime}}^2\right)}\left\{\delta^{s^{\prime} s} \delta^{r^{\prime} r}\left(1+\frac{|\vec{q}|^2}{4 m_\chi m_e}-\frac{|\vec{q}|^2}{8 \mu_{\chi e}^2}-\frac{|\vec{q}|^2}{8 \mu_{\chi e} m_e}\right)\right. \\
-i\left(\frac{1}{2 m_\chi} \delta^{r^{\prime} r}\left(\vec{S}_\chi^{s^{\prime} s} \times \vec{q}\right)+\frac{1}{m_e} \delta^{s^{\prime} s}\left(\vec{S}_e^{r^{\prime} r} \times \vec{q}\right)\right) \cdot \vec{v}_{e l}^{\perp}-\frac{1}{m_\chi m_e}\left(\vec{S}_\chi^{s^{\prime} s} \times \vec{q}\right) \cdot\left(\vec{S}_e^{r^{\prime} r} \times \vec{q}\right) \\
\left.+\left(\frac{\vec{k}}{m_e}\right) \cdot\left[\delta^{s^{\prime} s} \delta^{r^{\prime} r}\left(-\vec{v}_{e l}^{\perp}-\frac{\vec{q}}{4 \mu_{\chi e}}\right)+\frac{i}{2}\left(-\frac{1}{m_e} \delta^{s^{\prime} s}\left(\vec{S}_e^{r^{\prime} r} \times \vec{q}\right)+\frac{1}{m_\chi} \delta^{r^{\prime} r}\left(\vec{S}_\chi^{s^{\prime} s} \times \vec{q}\right)\right)\right]\right\}.
\end{gathered}
$$
%%%%%%%%%%%%%%%%%%%%%%%%%%%%%%%%%%%%%%%%%%%%%%%%%%%%%%%%%%%%%%
Neglecting the atomic electrons momentum $\vec{k}=0$

$$
\begin{gathered}
i \mathcal{M}=i \frac{4 m_\chi m_e g_\chi \epsilon e}{\left(|\vec{q}|^2+m_{Z^{\prime}}^2\right)}\left\{\delta^{s^{\prime} s} \delta^{r^{\prime} r}\left(1+\frac{|\vec{q}^2|}{4 m_\chi m_e}-\frac{|\vec{q}|^2}{8 \mu_{\chi e}^2}-\frac{|\vec{q}|^2}{8 \mu_{\chi e} m_e}\right)\right. \\
\left.-i\left(\frac{1}{2 m_\chi} \delta^{r^{\prime} r}\left(\vec{S}_\chi^{s^{\prime} s} \times \vec{q}\right)+\frac{1}{m_e} \delta^{s^{\prime} s}\left(\vec{S}_e^{r^{\prime} r} \times \vec{q}\right)\right) \cdot \vec{v}-\frac{1}{m_\chi m_e}\left(\vec{S}_\chi^{s^{\prime} s} \times \vec{q}\right) \cdot\left(\vec{S}_e^{r^{\prime} r} \times \vec{q}\right)\right\}
\end{gathered}
$$

Let's define a spin-dependent part

\begin{equation}
\mathcal{M}_{\rm SI}=\delta^{s^{\prime} s} \delta^{r^{\prime} r}\left(1+\frac{|\vec{q}|^2}{4 m_\chi m_e}-\frac{|\vec{q}|^2}{8 \mu_{\chi e}^2}-\frac{|\vec{q}|^2}{8 \mu_{\chi e} m_e}\right) 
\end{equation}
and a spin-dependent part
\begin{equation}
\mathcal{M}_{\rm SD}=-i\left(\frac{1}{2 m_\chi} \delta^{r^{\prime} r}\left(\vec{S}_\chi^{s^{\prime} s} \times \vec{q}\right)+\frac{1}{m_e} \delta^{s^{\prime} s}\left(\vec{S}_e^{r^{\prime} r} \times \vec{q}\right)\right) \cdot \vec{v}-\frac{1}{m_\chi m_e}\left(\vec{S}_\chi^{s^{\prime} s} \times \vec{q}\right) \cdot\left(\vec{S}_e^{r^{\prime} r} \times \vec{q}\right)
\end{equation}

Such that
\begin{equation}
i \mathcal{M}=i \frac{4 m_\chi m_e g_\chi \epsilon e}{\left(|\vec{q}|^2+m_{Z^{\prime}}^2\right)}\left\{\mathcal{M}_{\rm SI}+\mathcal{M}_{\rm SD}\right\}.
\end{equation}

\printbibliography

\end{document}